\documentclass[prb,twocolumn,superscriptaddress]{revtex4}
\usepackage{graphicx}
\usepackage{amsmath}
\usepackage{amssymb}
\usepackage{xspace}

\def\LCMO{$\mathrm{La_{0.50}Ca_{0.50}MnO_3}$\xspace}
\def\LCMOx{$\mathrm{La_{1-x}Ca_{x}MnO_3}$\xspace}
\def\PCMOx{$\mathrm{Pr_{1-x}Ca_{x}MnO_3}$\xspace}
\def\LCMOfiftwo{$\mathrm{La_{0.48}Ca_{0.52}MnO_3}$\xspace}
\def\PCMOfiftwo{$\mathrm{Pr_{0.48}Ca_{0.52}MnO_3}$\xspace}
\def\NGO{NdGaO$_3$\xspace}
\def\STO{SrTiO$_3$\xspace}

\def\Vec#1{\ensuremath{\mathbf{#1}}\xspace}

\begin{document}

\title{Very weak electron-phonon coupling and strong strain coupling in manganites}
\author{S. Cox}
\affiliation{National High Magnetic Field Laboratory, Ms-E536, Los Alamos National Laboratory, Los Alamos, New Mexico, 87545, USA}
\affiliation{Department of Materials Science and Metallurgy, University of Cambridge, U.K.}
\author{J.C. Loudon}
\affiliation{Cornell Center for Materials Research, Cornell University, New York, 14853, USA}
\author{A.J. Williams}
\affiliation{Centre for Science at Extreme Conditions, University of Edinburgh, Edinburgh, EH9 3JZ, UK}
\author{J.P. Attfield}
\affiliation{Centre for Science at Extreme Conditions, University of Edinburgh, Edinburgh, EH9 3JZ, UK}
\author{J. Singleton}
\affiliation{National High Magnetic Field Laboratory, Ms-E536, Los Alamos, New Mexico, 87545, USA}
\author{P.A. Midgley}
\affiliation{Department of Materials Science and Metallurgy, University of Cambridge, U.K.}
\author{N.D. Mathur}
\affiliation{Department of Materials Science and Metallurgy, University of Cambridge, U.K.}

\begin{abstract}

We describe transmission electron microscopy experiments that demonstrate the validity of the charge density wave (CDW) Landau theory 
in describing the so-called stripe phase of the manganites and that permit quantitative estimates of some of the 
theoretical parameters that describe this state.
In polycrystalline \PCMOfiftwo a lock-in to $q/a^*=0.5$ in a sample with $x>0.5$
has been observed for the first time.  Such a lock-in has been predicted as a key part of the Landau CDW theory of the stripe phase.
Thus it is possible to constrain the size of the electron-phonon coupling in the CDW Landau theory 
to between 0.08\% and 0.50\% of the electron-electron coupling term.
In the thin film samples, films of the same thickness grown on two different substrates exhibited different 
wavevectors.  The different strains present in the films on the two substrates can be related to the 
wavevector observed via Landau theory.  It is demonstrated that the 
the elastic term which
favours an incommensurate modulation has a similar size to the coupling between the strain and the wavevector, 
meaning that the coupling of strain to the superstructure is unexpectedly strong.

\end{abstract}

\maketitle

\section{Introduction}
The properties of the manganites comprise a zoo of delicately balanced phases in 
which changes in temperature, magnetic field, chemical composition and  strain 
(among other parameters) yield a rich tapestry of phase coexistence~\cite{neil_phystoday, neil_ssc}.
The stripe phase which forms at low temperatures on the insulating (resistivity decreases with increasing temperature) side of the manganite phase diagram was
long thought to be driven by strong electron-phonon coupling~\cite{Goodenough,CO1}, since the materials are insulating at 
all temperatures.  However, recent experimental work has led to the conclusion that the stripe phase 
is actually a charge density wave (CDW) subject to a high level of disorder~\cite{mangares}.  The CDW model 
of the stripe phase was described theoretically
using a Landau theory which successfully reproduced the observed variation of the wavevector $q$ with composition~\cite{milward}
($q/a^*=1-x$), where $a^*$ is the reciprocal lattice vector.  
This theory made the prediction that for $x$ slightly greater than 0.5
there should be a lock-in of the wavevector to $q/a^*=0.5$, though no such lock-in had ever been observed.  Here
we report the first observation of such a lock-in, in \PCMOfiftwo.  In addition, the values of the wavevector that we observe
can be used to constrain the Landau theory parameter for the electron-phonon coupling relative to the electron-electron coupling
to the range 0.08\%-0.50\%, suggesting that the electron-phonon coupling in the manganites is extremely weak.
This supports the model of a prototypical CDW, which is only weakly tied to the lattice, for the manganite stripe phase.

Landau theory also predicts~\cite{theory_strain, maria_strain} that it should be possible to tune the low temperature superstructure of the manganites 
by altering the strain in a small area of a thin film.  
Thus far it has not been possible to measure the changes in strain~\cite{me_strain},
and so it has not been possible to quantitatively link such changes in property
to theoretical work.
Here for the first time we measure the properties of the low temperature superstructure
in different known strain states using thin films of \LCMO, and provide a quantitative analysis of the 
results, allowing us to compare the size of the strain coupling and electron-electron coupling via the Landau theory.

For the thin film experiments, the composition \LCMO was chosen since it is extremely 
well characterised, and is the only manganite for which the stripe phase has been observed 
in a thin film before~\cite{me_strain}.
In polycrystalline \LCMO,  $q/a^*$ exhibits values between 0.46 and 
0.50 at 90~K (the intergranular variation is up to 9\%; the 
intragranular variation is less than 1\%)~\cite{philmag, CO1, chen_comm_incomm}.
Below the  N\'{e}el transition
temperature $T_\mathrm{N} \simeq 135$~K (on cooling)~\cite{chen_comm_incomm}
the positions of the superstructure reflections appear to stabilise, and it is assumed that in the absence of 
extrinsic factors $q/a^*$ would take the value 0.5.  The
superstructure persists up to the Curie temperature of $T_\mathrm{C} \simeq 220$~K, 
and for $T_\mathrm{N} < T < T_\mathrm{C}$, $q/a^*$
is hysteretic and incommensurate~\cite{chen_comm_incomm}.

Polycrystalline \LCMOfiftwo shows similar behaviour to \LCMO, except that $q/a^*$ does not lock in to 
0.5 in any grains (low temperature values lie between 0.43 and 0.475), and the variation of $q/a^*$ with temperature does not show a hysteresis loop~\cite{philmag}.
The compound \PCMOfiftwo was investigated here because it is predicted to have stronger electron-phonon coupling than \LCMOfiftwo, 
since the Pr ion is smaller than the La ion.
Stronger electron-phonon coupling should be associated with a stronger bias towards the stripe phase
as opposed to the ferromagnetic (FM) phase.  This is because $q/a^*=0.5$ is always observed if the stripe phase exists 
below $x=0.5$~\cite{pcmo_phase1, pcmo_phase2}; stronger electron-phonon coupling means a larger energy gain when the superstructure locks into the lattice. 
This effect can be observed in the phase diagrams for the two materials,
in which the \PCMOx stripe phase region extends down to $x$=0.3, whereas for \LCMOx it extends only to $x$=0.5~\cite{pcmo_phase1, pcmo_phase2, CMRoxidesb}.
Thus it should be more energetically favourable for the 
superstructure to lock into the lattice in  Pr$_{0.48}$Ca$_{0.52}$MnO$_3$ than in  La$_{0.48}$Ca$_{0.52}$MnO$_3$. 

The paper is organised as follows: Section~\ref{exptaldetails} gives a description of the sample preparation and experimental setup, 
Section~\ref{exptalresults} describes 
the results of the experiments on thin film \LCMO and polycrystalline \PCMOfiftwo, whilst Section~\ref{discussion} discusses  
these results in the context of the Landau theory for manganites. A summary is given in Section~\ref{conclusions}.

\section{Experimental details}
\label{exptaldetails}
Films were grown by pulsed laser deposition on 
\NGO and \STO substrates~\cite{me_strain}.
Three films were grown on \NGO, with thicknesses of $44\pm2$~nm, 
$80\pm3$~nm and $123\pm3$~nm as measured by high resolution X-ray diffraction.  One film of 
thickness $44\pm2$~nm was grown on \STO.  The 44~nm thickness is sufficiently
low to preserve cube-on-cube epitaxy.  
Magnetisation was measured using a vibrating sample magnetometer, with measurements being made for
the films of 44~nm, 80~nm and 123~nm grown on \NGO, and a bare substrate.  The measurements for the films were 
obtained by subtracting the substrate measurement from the thin film measurement, after correcting for the 
different volumes.
The polycrystalline sample of \PCMOfiftwo was prepared as described in~\cite{dirty_peierls}.
The macroscopic stoichiometry of the polycrystalline samples 
is accurate to within 0.1\%.

All the samples were prepared for transmission electron microscopy by conventional grinding and dimpling, and were thinned to electron transparency
($\sim$100 nm) by argon ion milling.
Measurements were taken in a Philips CM30 transmission electron microscope (TEM) with a 90~K liquid nitrogen stage.  
Each diffraction pattern was taken over a timescale of seconds.  Conventional imaging of the polycrystalline sample indicated that the 
grain size was $\sim$2~$\mu$m.

\begin{figure}
\begin{centering}
\includegraphics[width=0.9\columnwidth]{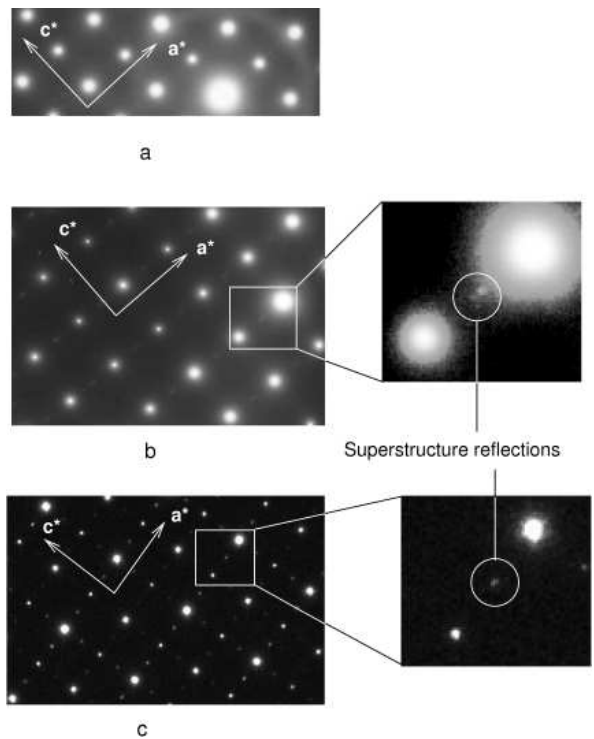}
\caption{Diffraction patterns for \LCMO thin film on \NGO and \STO. (a) Shows a room temperature diffraction pattern 
(for a film on \NGO) in which
no superstructure reflections appear and (b) shows a diffraction pattern for a film on \NGO taken at 90~K in which the superstructure
reflections are clearly visible. (c)
Shows a diffraction pattern taken at 90~K from a thin film grown on \STO.
\label{congo}}
\end{centering}
\end{figure}

\begin{figure}
\begin{centering}
\includegraphics[width=0.8\columnwidth]{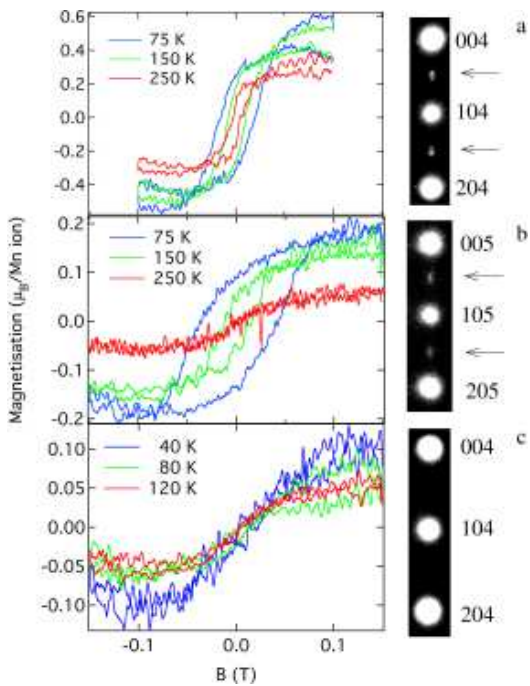}
\caption{(color online) $M-H$ loops at various temperatures for \LCMO thin films of (a) 44~nm (b) 80~nm and (c) 123~nm thickness on \NGO.  The 
diffraction patterns at 90~K for each film are also displayed, with the superstructure reflections being visible for (a) and (b).  The three diffraction
patterns have been enhanced to the same degree by high pass filtering and saturation.
\label{mhloops}}
\end{centering}
\end{figure}

\section{Experimental results}
\label{exptalresults}
\subsection{La$_{0.50}$Ca$_{0.50}$MnO$_3$ thin films}
The films grown on \NGO gave a uniaxial superstructure i.e. at 90~K the superstructure reflections appear along only one axis
(see Figures~\ref{congo}b).
The strength of the magnetisation and of the superstructure reflections were monitored for different film 
thicknesses (see Figure~\ref{mhloops}).  
The magnetization increased with decreasing temperature in all samples, as has previously been observed~\cite{xiong_film}.
It can be seen that the magnetization at low temperatures 
decreases in amplitude with increasing film thickness (from 0.5~$\mu_\mathrm{B}$/Mn~ion at 44~nm to 0.2~$\mu_\mathrm{B}$/Mn~ion at 80~nm to 0.1~$\mu_\mathrm{B}$/Mn~ion at 123~nm).
In addition, the temperature at which magnetic hysteresis appears decreases from 250~K in the 44~nm thick film to 175~K in the 80~nm
thick film, while no hysteresis appears in the 123~nm thick film.  However, the superstructure is strongest and most distinct
for the 44~nm thick film, being much fainter in the 80~nm thick film and not detectable in the 123~nm thick film.
Thus both the superstructure and the magnetisation become stronger with decreasing film thickness.
This has two implications.  Firstly, since the strain would be expected to decrease with increasing film thickness, 
it suggests that decreasing the level of strain in the film suppresses the superstructure and that in fact in the absence
of any strain the superstructure cannot exist.  Secondly, it suggests that rather than being in competition, the FM
and stripe phases coexist; if the stripe phase is destroyed, a FM phase is not formed, but rather a paramagnetic insulator (PMI).
This is supported by recent work which has found that for thin films 
there is an area 
of the phase diagram ($x>0.42$, limited by the maximum $x$ of 0.45 measured) for which the low temperature phase is a 
PMI, i.e. there are no signs of the FM or stripe phases~\cite{Diana_film}.  

\begin{figure}
\begin{centering}
\includegraphics[width=1.0\columnwidth]{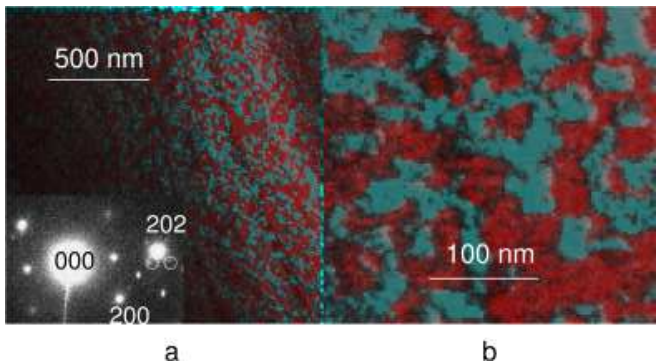}
\caption{Twins of two different orientations of the low temperature superstructure are imaged by taking two dark field images, one 
from each of the circled reflections in the diffraction pattern shown as the insert of (a), and superimposing them. (As before, only
one of the twins has been indexed).
One dark field image has been tinted red and the other turquoise, 
and the lack of overlap between the two colours indicates that the two superstructure orientations do not coexist.  The structure of the 
regions can be seen more clearly in (b). \label{overlaid}}
\end{centering}
\end{figure}

By contrast with films grown on \NGO, films grown on \STO gave diffraction patterns in which superstructure 
reflections appeared along both axes (see Figure~\ref{congo}).  By taking dark field images using superstructure
reflections from each of the two directions (see Figure~\ref{overlaid}), it was shown that the two superstructure 
orientations did not coexist.  Instead, the images revealed small, complementary regions of the different ordering
orientations for the superstructure.  The region of each twin extends over a length scale of 50-100~nm.
The twinning occurs for films grown on \STO but not on \NGO because of the different symmetries of the substrate:
\NGO has an orthorhombic structure, as does \LCMO, and so it is always most favourable for the \textbf{a} axis of 
\LCMO to align with the \textbf{a} axis of \NGO.  However, \STO is cubic, and thus it is equally favourable for the 
\textbf{a} axis of \LCMO to align along either the \textbf{a} or \textbf{c} axes of \STO.

A further difference between the properties of the \LCMO superstructure in the two types of film was the value of the 
wavevector.  
Diffractions patterns of a large area were taken using a selected area aperture of 500~nm diameter, as measured in the sample plane.  
The selected area diffraction 
patterns were analysed using software which measured the position of many superstructure reflections in one pattern~\cite{TEM_comp}.
The film on \NGO had a wavevector of $q/a^*=0.475$ in regions away from the edges, whereas the film grown on \STO had a wavevector of 
$q/a^*=0.50$.

\subsection{TEM measurements of Pr$_{0.48}$Ca$_{0.52}$MnO$_3$}
\begin{figure}
\begin{centering}
\includegraphics[width=0.49\textwidth]{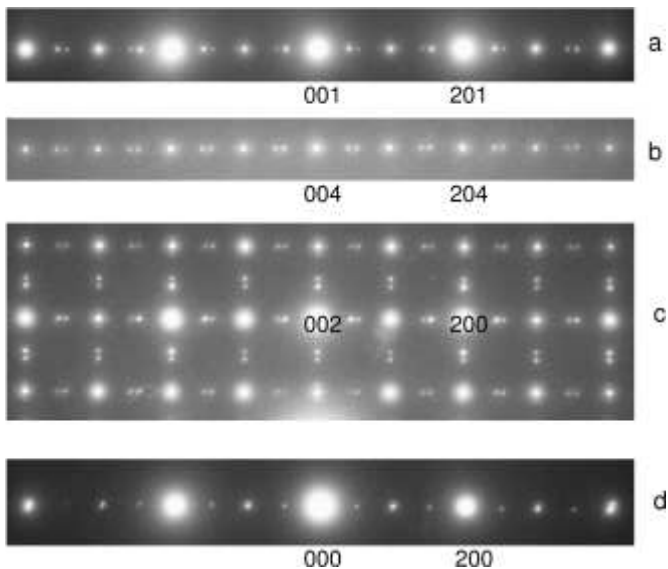}
\caption{Selected area diffraction patterns taken with a 500~nm aperture from different grains of the Pr$_{0.48}$Ca$_{0.52}$MnO$_3$ sample.  
Pattern (c) shows twinning.  Patterns a, b and c show wavevectors with $q/a^*$ between 0.445 and 0.45, whereas pattern 
d shows $q/a^*=0.5$.\label{SADpcmo}}
\end{centering}
\end{figure}

We now turn to the TEM measurements of \PCMOfiftwo.
Selected area diffraction patterns were taken using an aperture of 500~nm diameter, as described above.
The size of the aperture was much smaller than the grain size (2~$\mu$m~\cite{philmag}).  Therefore the 
diffraction patterns measure intrinsic proerties of the grains, including strain.
In all patterns $\Vec{q}$ was found to be essentially parallel to $\Vec{a^*}$.
Three diffraction patterns (Figure~\ref{SADpcmo}a,b,c) have values of $q/a^*$ between 0.445 and 0.450, 
and the fourth shows $q/a^*=0.5$ (Figure~\ref{SADpcmo}d).  Therefore in \PCMOfiftwo the superstructure will lock into the lattice
under certain conditions, supporting the idea that the electron-lattice coupling is stronger in this compound
than in La$_{0.48}$Ca$_{0.52}$MnO$_3$. 
This is the first observation of a lock-in to $q/a^*=0.5$ for $x>0.5$ in a manganite.

The measurements of the wavevector in the grains in which no lock-in occurred are 
within the range one would expect from previous measurements of \LCMOfiftwo.  It should be noted that the nominal value of 
$q/a^*=1-x$ is never actually observed, in fact $q/a^*<1-x$.  This deviation and the variation of wavevector from grain to grain 
is traditionally ascribed to different levels of strain in the grains.
This can be described in the context of a Ginzberg-Landau theory~\cite{Landau, milward}. 

\begin{figure}
\begin{centering}
\includegraphics[width=0.5\textwidth]{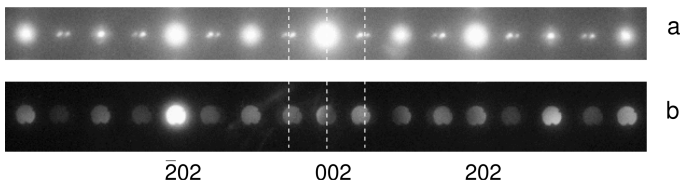}
\caption{(a) Selected area diffraction and (b) convergent beam electron diffraction patterns from the same region of a Pr$_{0.48}$Ca$_{0.52}$MnO$_3$ grain.  The wavevectors are the same within 
experimental error.  The dotted line is at $q/a^*=0.5$.\label{CBEDpcmo}}
\end{centering}
\end{figure}

Convergent beam electron diffraction patterns were obtained using a converged beam with a full width at half maximum of 3.6 nm, 
which corresponds to 6.7 room temperature 
unit cells in Pr$_{0.48}$Ca$_{0.52}$MnO$_3$.  
From the composition $x=0.52$ in a one dimensional model of charge order using Mn$^{3+}$ and Mn$^{4+}$ one would expect alternating  Mn$^{3+}$ and Mn$^{4+}$, 
with an extra Mn$^{4+}$ every 13.6 unit cells on average~\cite{philmag}.  This would lead to a value of $q/a^*=0.5$ being recorded in most measurements.  
However, the convergent beam electron diffraction patterns clearly showed a periodicity 
equivalent within error to the one extracted from the corresponding selected area diffraction pattern (Figure~\ref{CBEDpcmo}).  
Thus the periodicity of the superstructure is uniform down to the level of a few unit cells in Pr$_{0.48}$Ca$_{0.52}$MnO$_3$.

\begin{table}
\begin{tabular}{|l|c|c|c|c|}
\hline
          &		$\eta_a$ at 90 K	&	    $\eta_c$ at 90 K      &$\eta_a$ at  300 K 	     &	  $\eta_c$ at 300 K    \\
\hline
\NGO        &   		-0.0048		&		0.0035  	&		0.0017  	 &  0.0119  \\
\STO         &   		0.0012		&	0.0066  	&  		0.0176  	 &  0.0157  \\
\hline
\end{tabular}
\caption{In-plane strains for \LCMO on NdGaO$_3$ and SrTiO$_3$ at 90~K and 300~K.  Positive 
numbers refer to the \LCMO parameter being smaller than the substrate parameter, so the film is stretched.
Data from~\cite{Radaelli1/2,ngo_temp,  sto2}.\label{mismatch}}
\end{table}

\section{Discussion}
\label{discussion}
As has been previously shown~\cite{me_strain}, the free energy of the superstructure in a Ginzberg-Landau theory is given by:
\begin{equation}
F = \frac{\xi^2}{2} (\nabla\phi - \delta)^2 + \frac{v}{n}\cos(n\phi) + c\eta\nabla\phi + \frac{1}{2}\kappa\eta^2 - \sigma\eta.
\label{eq:energy}
\end{equation}
where first term is the elastic term that favors incommensurate modulation and $\delta$ is the deviation of $q/a^*$ in the absence of strain coupling.
The second term is the Umklapp term that favors commensurability, where $n$ is the commensurability and $v$ determines the strength of the effect.
In this case $n=4$ since the periodicity of the low temperature superstructure is four times the room temperature pseudocubic cell~\cite{maria_strain}.
The third term couples $\eta$ and $\nabla\phi$ with strength $c$, the fourth term is the strain energy density in terms of the bulk elastic modulus 
$\kappa$ and the fifth term gives the elastic enery due to the stress $\sigma$ on the film from the substrate.
For ease of reference, these parameters are defined in Table~\ref{parameters}, which also summarises the values obtained in the experiments 
described in this paper.

Minimising $F$ in the plane-wave limit ($\nabla\phi$=constant and $\nabla\eta$=0), we find:
\begin{equation}
\nabla\phi = \frac{\delta - \frac{c\sigma}{\kappa\xi^2}}{1-\frac{c^2}{\kappa\xi^2}}.
\end{equation}

To calculate the approximate levels of stress in each film, the strain in the $a$ and $c$ directions must first be found.
Let SrTiO$_3$ be substrate 1 and NdGaO$_3$ be substrate 2. 
Calculating the strain using the mismatch of the $a$ and $c$ lattice parameters for the substrate and \LCMO, we obtain 
$\eta_{1a}=0.012$, $\eta_{1c}=0.0066$, $\eta_{2a}=-0.0048$ and $\eta_{2c}=0.0035$ (see Table~\ref{mismatch}). 
These values can be used to calculate the level of strain in the films, using $\sigma_a = E(\eta_a + \nu\eta_c)/(1-\nu^2)$, where $E$ is 
the Young's modulus of the material.
Using approximate value of the bulk modulus $\kappa$ (135~GPa) and shear modulus (31~GPa)~\cite{maria_strain}, 
we obtain a Young's modulus of 86~GPa and a Poisson's ratio ($\nu$) of 0.39, giving
$\sigma_{1a}=0.00405E$ and $\sigma_{2a}=0.0172E$.

Considering the results for the film on \STO, with $\nabla\phi_1=0$:
\begin{equation}
\frac{\sigma_{1a} c}{\kappa\xi^2}=\delta.
\label{eq:ca}
\end{equation}

To obtain an approximate value of $\delta$ we consider the results for polycrystalline 
samples, and assume that the extraneous effects which render $\delta$ finite have different values in the different grains, 
and that in the thin films these effects are around the average level that they are in the different polycrystalline grains.  
Therefore, $\delta=0.0124a^*$ (the average $\delta$ for observations in polycrystalline \LCMO samples).

Substituting into equation (\ref{eq:ca}) gives $c/\xi^2 = -1.1$ (Table~\ref{parameters}).
So surprisingly, the strain coupling term is at a similar level to the 
elastic term that favours an incommensurate modulation.

Now we consider the film on \NGO;
substituting the values of $c/\xi^2$ and $\nabla\phi_{2a}=-0.025a^*$ into equation~\ref{eq:energy} we find:  
\begin{equation}
\frac{c^2}{\kappa\xi^2} = 1 - \frac{\delta}{\nabla\phi_{2a}}\left(1-\frac{\eta_1}{\eta_2}\right)
\label{eq:ratio}
\end{equation}
Substituting in gives $c/\kappa=-0.3$.  So the strain energy density term is larger than the coupling term and the 
elastic incommensurate term.

\begin{table*}
\begin{tabular*}{2\columnwidth}{@{\extracolsep{\fill}}|p{0.6in}|p{2.3in}|p{1.2in}|p{1.0in}|p{1.4in}|}
\hline
Parameter          &		Meaning	&	    \multicolumn{2}{c|}{Value/Constraint}      & Measured or derived 	        \\
                   &                &        \LCMOfiftwo    &    \PCMOfiftwo           &                               \\
\hline
$\eta$             &  Strain due to substrate			& See Table I		  	&	---	  	 &  Derived from substrate/lattice mismatch  \\
$\xi$         &  Electron-electron coupling        &  ---                     &    ---        &  Unknown, but see $v$ and $V$ below             \\
$\delta$          &  Deviation of $q$ from $0.5a^*$	in the absence of strain from a substrate	& positive - estimated value $0.0124a^*$  	&  positive 	 &  Inferred from literature  \\
$\delta_\mathrm{s}$      &  Deviation of $q$ from $0.5a^*$	in the absence of lock in term 	& 0.03$a^*$---0.07$a^*$	(from~\cite{philmag})  	&  	0.05$a^*$---0.02$a^*$	  	 &  Measured from polycrystalline results  \\
$c$                &  Coupling between strain and superstructure 				&	$c/\xi^2$=-1.1, $c/\kappa$=-0.3  	&  ---		  	 &  Derived in this paper  \\
$v$                &  Coupling between commensurate periodicity and superstructure 	&	$2v/n\xi^2<0.0009$, or $v/\xi^2<0.0018$  	&   $4\times10^{-4}<2v/n\xi^2<0.0025$, or $8\times10^{-4}<v/\xi^2<0.005$ 		  	 &  Derived in this paper  \\
$n$           &   Commensurability				&	4  	&  4		 & Inferred from superstructure periodicity   \\
$\sigma$      &   Stress due to substrate				&	0.00405$E$ on SrTiO$_3$, 0.0172$E$ on NdGaO$_3$  	& --- 		  	 &   Derived from $\eta$ and $E$ \\
$\kappa$      &  Bulk modulus 				&	135 GPa  	&---  		  	 &  Taken from~\cite{maria_strain}   \\
$S$           &  Shear modulus 				&	31 GPa  	& --- 		  	 &  Taken from~\cite{maria_strain}  \\
$E$           &  Young's modulus 				& 86 GPa	  	&---  		  	 &  Derived from $S$ and $\kappa$  \\
$V$           &  =(2$v/n\xi^2$)  = ($v/2\xi^2$) where $v/\xi^2$ is the  ratio between strength of electron-phonon coupling and electron-electron coupling	&	 $<0.0009$ 	&  	0.0004---0.0025, lower values have higher probability	  	 &  Derived in this paper  \\
\hline
\end{tabular*}
\caption{Table showing the meaning and values of the parameters used and derived in this paper.{\label{parameters}}}
\end{table*}

\begin{figure}
\begin{centering}
\includegraphics[width=0.40\textwidth]{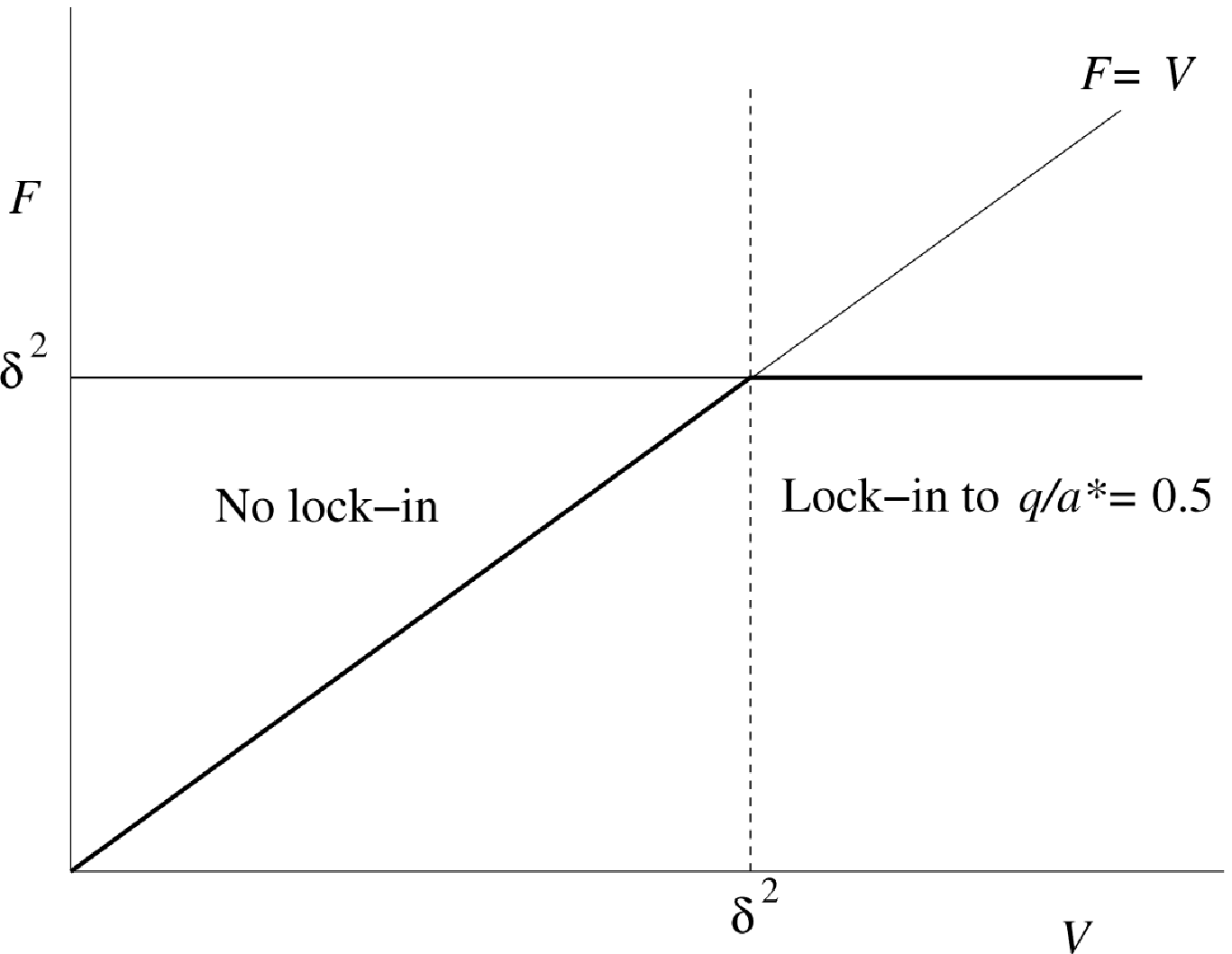}
\caption{Demonstrates the relations between $V$ and $F$ which leads to a lock-in for $V<\delta^2$ and to no lock-in for $V>\delta^2$.\label{lockin}}
\end{centering}
\end{figure}

The Ginzberg-Landau theory was used to investigate which sets of parameters could give rise to the observed wavevector values in polycrystalline \PCMOfiftwo.
Since we cannot quantify the levels of strain in the different grains, we take $\delta_\mathrm{s}$ to include the deviation from $q/a^*=0.5$
which is due to strain.  Then the terms in the free energy which can vary are given by:
\begin{equation}
F=(\nabla\phi - \delta_\mathrm{s})^2 + V \cos(n\phi)
\end{equation}
where $V=2v/n\xi^2=v/2\xi^2$.  Here, $v/\xi^2$ is the ratio between the strength of the electron-phonon coupling and the electron-electron coupling.
The ground state can
be found by minimising this quantity with respect to $\phi$, with the boundary conditions that at $\phi=0$, $\nabla\phi=\delta_\mathrm{s} + t$, where
$t<<\delta_\mathrm{s}$.  In the limit of small $t$, this gives $F=V$.
The free energy of this state can be calculated and compared to the energy of the lock-in state, which has $\phi=\pi/8$, $\nabla\phi=0$, and thus
$F=\delta^2$.  So if the energy of the lock-in state is lower, the superstructure will lock in and $q/a^*=0.5$.

Since the wavevector does not lock into the lattice for $\delta_\mathrm{s}=0.05$, it follows that $V$ must be smaller than $\delta_\mathrm{s}^2$, since above this level a lock-in
should occur.  However, we know that a lock-in does occur in one grain.  Therefore in that grain $\delta$ is taken to be the smallest value previously observed
in a manganite with $x=0.52$, which is $\delta_\mathrm{s}=0.02$.  Therefore the lock-in provides a lower bound for $V$, since $V$ must be larger than $\delta_\mathrm{s}^2$ in 
this case to allow a lock-in to occur.  So $0.0025<V<0.0004$ for \PCMOfiftwo (Table~\ref{parameters}).  This method can also be used to provide an upper bound for $V$
in \LCMOfiftwo, since no lock-in is observed when $q/a^*=0.03$, giving $V < 0.0009$ (Table~\ref{parameters}).

\begin{figure}
\begin{centering}
\includegraphics[width=0.4\textwidth]{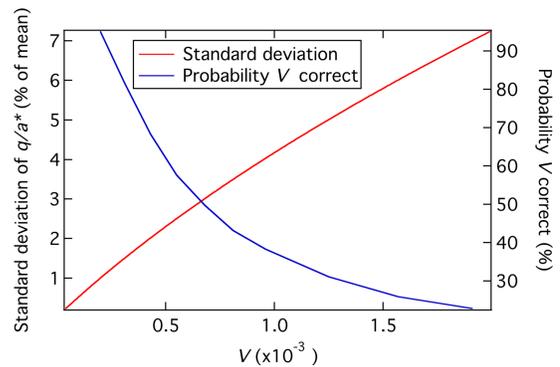}
\caption{Variation of the standard deviation of $q/a^*$ with $V$ (red) and the probability that a given value of $V$ could produce the observed 
result (blue).\label{eta_var}}
\end{centering}
\end{figure}

The significance of the value of the wavevector being the same at lengthscales of 3.6~nm and 500~nm was then investigated.
The equations shown above were used to calculate the values of $\phi$ and $\nabla\phi$ for an array of 925 room temperature 
unit cells (equivalent to the diameter of the 500~nm aperture).  From this array 90 non-overlapping regions of 6.7 unit cells were chosen.
The standard deviations of the average wavevectors of these regions were found for different values of $V$.  The results are shown in Figure~\ref{eta_var} displayed as a percentage of 
the 925 cell average.  Given the 1\% standard deviation in the measurement of the wavevector for the selected area diffraction and convergent beam electron diffraction results, we wanted to 
know the probability of measuring a large and small scale value so close to each other for each value of $V$.  We assume that both 
the measurement of the wavevector and the simulation of the areas of 6.7 unit cells have Gaussian distributions.  The probability of the 
convergent beam electron diffraction result being within 4\% of the selected area diffraction result experimentally is then 95\%.  Then taking the distribution of the simulation, the probability of 
the small-scale results being within 4\% of the large-scale result can be calculated from the standard deviation at each value of $V$.
This indicates the probability of obtaining the result we did for each value $V$.  As can be seen, the probability decreases 
with increasing $V$, meaning that the values of $V$ close to 0.0004 are more likely to be correct.
All parameters are summarised in Table~\ref{parameters}.

\section{Conclusions}
\label{conclusions}
The CDW Landau theory of the manganites predicts that a lock-in should occur in the manganite superstructure when $x>0.5$, for some range of $x$
close to 0.5.  In this paper we report the first observation of such a lock-in.  
It is also expected that the electron-phonon coupling parameter should be small relative to the electron-electron
coupling parameter.  Based on our data, we have used Landau theory to constain the value of the electron-phonon coupling relative to the 
electron-electron coupling to between 0.08\% and 0.50\%, with the results indicating that the lower end of this scale 
has a higher probability.
In addition, we have found that the coupling of the strain to the superstructure in the stripe phase is of the same magnitude as
the electron-electron coupling, indicating that the coupling of the superstructure to strain is unexpectedly strong.  
This is particularly interesting since our results also suggest that the stripe phase cannot exist in the absence of strain.
Therefore the properties of the manganite CDW can now be quantified to a reasonable extent (Table~\ref{parameters}).  Our results 
also raise the possibility of novel uses of strain to manipulate the CDW, via the strong coupling of 
strain to the CDW and the possibility of destroying the CDW by releasing the strain.

\section{Acknowledgements}
We thank P.B. Littlewood for helpful comments.
This work is supported by DoE grant LDRD-DR 20070013 and by the EPSRC.  
Work at the NHMFL is performed under the auspices of the NSF, DoE and the State of Florida.  Work at Cambridge was funded by the UK EPSRC and the 
Royal Society. S. Cox acknowledges support from the Seaborg Institute.

\end{document}